\documentclass[aps,pra,twocolumn]{revtex4}
\usepackage{amssymb}
\usepackage{graphicx}
\usepackage{dcolumn}
\usepackage{bm}

\begin{document}

\title{Ultracold two-component Fermi gases with a magnetic field gradient
near a Feshbach resonance}
\author{Hongwei Xiong$^{1,2,3}$, Shujuan Liu$^{1,2,3}$, Weiping Zhang$^{4}$,
Mingsheng Zhan$^{1,2}$}

\address{$^{1}$State Key Laboratory of Magnetic Resonance and
Atomic and Molecular Physics, Wuhan Institute of Physics and
Mathematics, Chinese Academy of Sciences, Wuhan 430071, P. R.
China}
\address{$^{2}$Center for Cold Atom Physics, Chinese Academy of Sciences, Wuhan 430071, P.
R. China}
\address{$^{3}$Graduate School of the Chinese Academy of Sciences,
Wuhan 430071, P. R. China}
\address{$^{4}$Department of Physics,
East China Normal University}

\begin{abstract}
We study theoretically the ultracold two-component Fermi gases when a
gradient magnetic field is used to tune the scattering length between atoms.
For $^{6}Li$ at the narrow resonance $B_{0}=543.25$ G, it is shown that the
gases would be in a coexistence of the regimes of BCS, BEC and unitarity
limit with the present experimental technique. In the case of thermal and
chemical equilibrium, we investigate the density distribution of the gases
and show that a double peak of the density distribution can give us a clear
evidence for the coexistence of BCS, BEC and unitarity limit.

\end{abstract}

\maketitle

With the remarkable development of the cooling technique for two-component
Fermi gases and Feshbach resonance, the recent experiments \cite%
{JOCHIM,JIN,MIT-mole} have finally realized molecular Bose-Einstein
condensation (BEC) and the condensation of fermionic atom pairs on the side
of attractive interaction has also been investigated by several experiments
\cite{JIN-fermion,MIT-fermion,GRIMM,SALOMON,THOMAS,GRIMM1,Cheng,HEAT}.
Together with intensive theoretical investigations \cite%
{LEGGETT,NOZI,STOOF,TIMM,OHASHI1,MILSTEIN,STAJ,WU,CARR,BRUUN,FALCO,PERALI,XIONGCROSS,HARA,HEISELBERG,CARLSON,HO1,HO,XIONGON,TORMA}
such as resonance superfluid \cite{TIMM,MILSTEIN} and universal behavior for
the gases with divergent scattering length \cite%
{HARA,HEISELBERG,CARLSON,HO1,HO,XIONGON}, these experimental advances
provide us a quite unique system which will even contribute largely to our
understanding on the mechanism of high-temperature superconductors.

For two-component Fermi gases, when the magnetic field is tuned so that the
energy of a quasibound molecular state in a closed channel matches the total
energy in an open channel, there is a magnetic-field Feshbach resonance \cite%
{HARA,FESHBACH-THE} which can tune the scattering length from positive to
negative over many orders of magnitude. On the side of repulsive interaction
(BEC side), there exists molecule which is a short-range fermionic atom
pairs. On the side of attractive interaction (BCS side), there would be a
superfluid behavior due to the atomic Cooper pairs at sufficient low
temperature. On resonance, the absolute value of the scattering length is
much larger than the average distance between atoms and one expects a
universal behavior for the system in the unitarity limit. In the present
experiments on the BCS-BEC crossover, a uniform magnetic field is used to
tune the scattering length through the magnetic-field Feshbach resonance. By
tuning the uniform magnetic field near the resonant magnetic field, several
experiments have investigated the BCS-BEC crossover such as condensate
fraction \cite{JIN-fermion,MIT-fermion}, collective excitation \cite%
{THOMAS,GRIMM1}, pairing gap \cite{Cheng}, and heat capacity \cite{HEAT}.

For ultracold two-component Fermi gases confined in an optical trap with
axial symmetry along the $z$ axis, in the present work, we will investigate
the unique property of the system when the magnetic field to tune the
scattering length has a gradient of $\alpha $. For $^{6}Li$ at the narrow
resonance $B_{0}=543.25$ \textrm{G}, our researches show that the gases can
be in a coexistence of the regimes of BCS, BEC and unitarity limit with
appropriate parameters. For the gases in thermal and chemical equilibrium,
when the pair size is much smaller than the cloud size of the system, the
density distribution of the gases is calculated based on the local density
approximation.

For Fermi gases with an equal incoherent mixture of the internal states $%
\left\vert 1\right\rangle $ and $\left\vert 2\right\rangle $, near a
Feshbach resonant magnetic field $B_{0}$, the scattering length is $a\left(
B\right) =Aa_{0}\left( 1-w/\left( B-B_{0}\right) \right) $ with $a_{0}$
being the Bohr radius. For the magnetic field in $x-$direction with gradient
$\alpha $ along $z-$direction, we have $\vec{B}=\left( B_{0}+\alpha z\right)
\vec{e}_{x}$. In this case, the scattering length becomes $z-$dependent and
takes the following form:
\begin{equation}
a\left( z\right) =Aa_{0}\left( 1-\frac{w}{\alpha z}\right) .
\end{equation}%
If the magnetic field gradient $\alpha $ is positive, when $z<0$ and $%
a\left( z\right) n^{1/3}<1$ ($n$ is the total density distribution of the
ultracold gases), there is repulsive interaction between fermionic atoms
which corresponds the BEC side, while when $z>0$ and $\left\vert a\left(
z\right) \right\vert n^{1/3}<1$ there is attractive interaction between
fermionic atoms which corresponds to the BCS side. In the regime determined
by $\left\vert a\left( z\right) \right\vert n^{1/3}>1$, the ultracold gases
are in the unitarity limit where the scattering length can be regarded to be
divergent and will not appear in the final result of a physical quantity
such as chemical potential. Thus, by using the gradient magnetic field and
appropriate parameters, there would be a coexistence of the gases in the
form of BCS, BEC and unitarity limit.

For the optical trap with angular frequencies being $\omega _{r}(=\omega
_{x}=\omega _{y})$ and $\omega _{z}$ in the radial and axial directions,
when the magnetic field with gradient $\alpha $ is applied to the ultracold
gases, the overall external potential for the fermionic atom in the internal
state $\left\vert i\right\rangle $ is
\begin{equation}
V_{\left\vert i\right\rangle }=\frac{1}{2}m\omega _{r}^{2}\left(
x^{2}+y^{2}\right) +\frac{1}{2}m\omega _{z}^{2}\left( z-\Delta z_{\left\vert
i\right\rangle }\right) ^{2}-\mu _{\left\vert i\right\rangle
}^{mag}B_{0}+\Delta V_{\left\vert i\right\rangle },
\end{equation}%
where $i=1,2$. Due to the presence of the magnetic field, for different
internal state $\left\vert i\right\rangle $, there is a different shift to
the optical trap which is determined by $\Delta z_{\left\vert i\right\rangle
}=\alpha \mu _{\left\vert i\right\rangle }^{mag}/m\omega _{z}^{2}$ with $\mu
_{\left\vert i\right\rangle }^{mag}$ being the magnetic moment of the
internal state $\left\vert i\right\rangle $. In the above expression, $%
\Delta V_{\left\vert i\right\rangle }=-\alpha ^{2}\left( \mu _{\left\vert
i\right\rangle }^{mag}\right) ^{2}/2m\omega _{z}^{2}$. One can also get the
overall external potential for a molecule
\begin{equation}
~{V_{mol}=m\omega _{r}^{2}\left( x^{2}+y^{2}\right) +m\omega _{z}^{2}z}%
^{2}-\mu _{m}^{mag}B,
\end{equation}%
where $\mu _{m}^{mag}$ is the magnetic moment of the molecule.

The regime of the gases in the unitarity limit can be roughly determined by $%
\left\vert a\left( z\right) \right\vert n^{1/3}>1$. Thus, the length scale
for the regime in the unitarity limit can be estimated as $2l_{z}^{res}$\
with $l_{z}^{res}=Awn^{1/3}a_{0}/\alpha $. Because there is a different
shift for the external potential of the internal state $\left\vert
1\right\rangle $\ and $\left\vert 2\right\rangle $, the difference of the
shift should be much smaller than $2l_{z}^{res}$\ so that one can omit
safely the effect of the shift $\Delta z_{\left\vert i\right\rangle }$.
After a simple calculation, this condition $\left\vert \Delta z_{\left\vert
1\right\rangle }-\Delta z_{\left\vert 2\right\rangle }\right\vert
<<2l_{z}^{res}$\ means that
\begin{equation}
Aw<<\frac{2\left( l_{z}^{res}\right) ^{2}m\omega _{z}^{2}}{%
n^{1/3}a_{0}\left\vert \mu _{\left\vert 1\right\rangle }^{mag}-\mu
_{\left\vert 2\right\rangle }^{mag}\right\vert }.  \label{condition}
\end{equation}

For the typical parameters that $\omega _{z}/2\pi \sim 100$ \textrm{Hz}, $%
n\sim 10^{13}$ \textrm{cm}$^{-3}$ and $l_{z}^{res}\sim 10$ $\mathrm{\mu m}$,
the condition given by Eq. (\ref{condition}) requests that $Aw$ is smaller
than $10$ \textrm{G}. This means that very narrow resonance for an element
is an appropriate choice to investigate the coexistence of the regimes of
BCS, BEC and unitarity limit. A careful investigation shows that the mixture
of $\left\vert 1\right\rangle \equiv \left\vert F=1/2,m_{F}=1/2\right\rangle
$ and $\left\vert 2\right\rangle \equiv \left\vert
F=1/2,m_{F}=-1/2\right\rangle $ for $^{6}Li$ at the narrow resonance located
at $B_{0}=543.25$ \textrm{G }\cite{HULET} can satisfy our request. For this
narrow resonance, $Aw$ is estimated as $6$ \textrm{G} \cite{HULET}. Thus, in
the present work, we will consider two-component Fermi gases of $^{6}Li$ at
the narrow resonance $B_{0}=543.25$ \textrm{G}.

For two-component Fermi gases of $^{6}Li$ confined in the optical trap, we
will investigate the density distribution of the gases by using the
condition that the system is in thermal and chemical equilibrium. Our
researches show that with appropriate choice of parameters and element for
the system, the coexistence of the gases in the form of BCS, BEC and
unitarity limit can be clearly shown through the density distribution. In
the present experiments, the gases can be cooled far below the critical
temperature of molecular BEC and Fermi temperature of the Fermi gases. Thus,
we will consider the system at zero temperature to give a clear
presentation. For the molecules on the BEC side, the pair size can be
regarded as the scattering length. For the atomic pairs in the unitarity
limit, the pair size can be estimated as the average distance between atoms
\cite{HO-science}. For the atomic Cooper pairs on the BCS side, one can
estimate the pair size based on the BCS theory. We have verified that for
the parameters used in the present work, the pair size is much smaller than
the cloud size. Thus, one can safely use the well-known local density
approximation to calculate the density distribution of the system.

To calculate the density distribution of the system, we firstly investigate
the chemical potential of the system. In the regime of molecular BEC ($%
a\left( z\right) n^{1/3}<1$) with density distribution $n_{m}^{BEC}$ and
molecular scattering length $a_{m}=0.6a$ \cite{PETROV}, the chemical
potential of the molecular BEC is given by
\begin{equation}
\mu _{m}^{BEC}=\frac{2\pi \hbar ^{2}a_{m}n_{m}^{BEC}}{m}+V_{mol}+\Delta \mu
B_{0},  \label{BEC-chemical}
\end{equation}%
where $\Delta \mu =\mu _{m}^{mag}-\Sigma _{i}\mu _{\left\vert i\right\rangle
}^{mag}$. The last term $\Delta \mu B_{0}$ in the above equation is due to
the energy of the bound state in the closed channel when there is no
magnetic field. After a simple calculation, one can get the following
expression for the chemical potential%
\begin{equation}
\mu _{m}^{BEC}=\frac{2\pi \hbar ^{2}a_{m}n_{m}^{BEC}}{m}+V_{mol}^{\prime
}+\varepsilon _{m},  \label{BEC-new-chemical}
\end{equation}%
where $\varepsilon _{m}=-E_{m}$ with $E_{m}=\hbar ^{2}/ma^{2}$ being the
binding energy of the molecule. $V_{mol}^{\prime }$ is given by%
\[
{V_{mol}^{\prime }=m\omega _{r}^{2}\left( x^{2}+y^{2}\right) +m\omega
_{z}^{2}\left( z-\Delta z_{mol}\right) ^{2}}
\]
\begin{equation}
~~~~-\Sigma _{i}\mu _{\left\vert i\right\rangle }^{mag}B_{0}+\Delta V_{mol},
\end{equation}%
where $\Delta z_{mol}=\alpha \left( \Sigma _{i}\mu _{\left\vert
i\right\rangle }^{mag}\right) /2m\omega _{z}^{2}$ and $\Delta
V_{mol}=-\alpha ^{2}\left( \Sigma _{i}\mu _{\left\vert i\right\rangle
}^{mag}\right) ^{2}/4m\omega _{z}^{2}$.

In this regime, the chemical potential of the Fermi gas in the internal
state $\left\vert i\right\rangle $ takes the form
\begin{equation}
\mu _{f\left\vert i\right\rangle }^{BEC}=\frac{\hbar ^{2}\left( 6\pi
^{2}\right) ^{2/3}}{2m}\left( n_{f\left\vert i\right\rangle }^{BEC}\right)
^{2/3}+g_{aa}n_{f\left\vert i\right\rangle
}^{BEC}+g_{am}n_{m}^{BEC}+V_{\left\vert i\right\rangle },
\end{equation}%
where $n_{f\left\vert i\right\rangle }^{BEC}$ is the density distribution of
the Fermi gas in the internal state $\left\vert i\right\rangle $. In the
above expression $g_{aa}=2\pi \hbar ^{2}a/m_{r}$ with $m_{r}=m/2$ being the
reduced mass. In addition, $g_{am}=0.9g_{aa}$ \cite{Chia} which is obtained
based on the atom-molecule scattering length $a_{am}=1.2a$ \cite{PETROV}.

In the regime of the unitarity limit where the absolute value of the
scattering length is much larger than the average distance between atoms, as
pointed out in \cite{XIONGON}, we assume that the gases is in the mixture of
Fermi gases and dimeric gas in chemical equilibrium. Based on the local
density approximation at zero temperature, the chemical potential of the
Fermi gas with density distribution $n_{f\left\vert i\right\rangle }^{UL}$
takes the form
\begin{equation}
\mu _{f\left\vert i\right\rangle }^{UL}=\left( 1+\beta _{1}\right) \frac{%
\hbar ^{2}\left( 6\pi ^{2}\right) ^{2/3}}{2m}\left( n_{f\left\vert
i\right\rangle }^{UL}\right) ^{2/3}+V_{\left\vert i\right\rangle },
\end{equation}%
where $\beta _{1}$ is firstly measured in \cite{HARA} and $\beta _{1}=-0.56$
based on a quantum Monte Carlo calculation \cite{CARLSON}. Omitting the
binding energy of the dimer in the unitarity limit, based on the
dimensionality analysis, the chemical potential of the dimeric gas with
density distribution $n_{d}^{UL}$ can be assumed as \cite{XIONGON}
\begin{equation}
\mu _{d}^{UL}=\left( 1+\beta _{2}\right) \frac{\hbar ^{2}\left( 6\pi
^{2}\right) ^{2/3}}{2\times 2m}\left( n_{d}^{UL}\right) ^{2/3}+V_{d}.
\end{equation}%
For the dimeric gas, $V_{d}$ takes the same form as $V_{mol}^{\prime }$ for
the molecular gas. Based on the condition of the chemical equilibrium $%
\Sigma _{i}\mu _{f\left\vert i\right\rangle }^{UL}=\mu _{d}^{UL}$, one can
get the ratio $n_{d}^{UL}/n_{f\left\vert i\right\rangle }^{UL}=\left[
4\left( 1+\beta _{1}\right) /\left( 1+\beta _{2}\right) \right] ^{3/2}$ on
resonance. From the experimental result $n_{d}^{UL}/n_{f\left\vert
i\right\rangle }^{UL}\approx 4$ in \cite{MIT-fermion}, $\beta _{2}$ is
estimated as $-0.3$.

In the regime of BCS, the chemical potential can be approximated as
\begin{equation}
\mu _{f\left\vert i\right\rangle }^{BCS}=\frac{\hbar ^{2}\left( 6\pi
^{2}\right) ^{2/3}}{2m}\left( n_{f\left\vert i\right\rangle }^{BCS}\right)
^{2/3}+V_{\left\vert i\right\rangle },  \label{BCS-chemical}
\end{equation}%
where $n_{f\left\vert i\right\rangle }^{BCS}$ is the density distribution of
the fermionic atoms on the BCS side.

In the case of thermal and chemical equilibrium for the system, the minimum
of the Gibbs free energy means that there is following important relation
for the chemical potential:
\begin{equation}
\mu _{m}^{BEC}=\Sigma _{i}\mu _{f\left\vert i\right\rangle }^{BEC}=\Sigma
_{i}\mu _{f\left\vert i\right\rangle }^{UL}=\mu _{d}^{UL}=\Sigma _{i}\mu
_{f\left\vert i\right\rangle }^{BCS}\equiv \mu .  \label{chem-condition}
\end{equation}%
To illustrate clearly the density distribution and evidence for an
experiment to show the coexistence of the regime of BCS, BEC and unitarity
limit, we use the parameters $\omega _{z}/2\pi =120$ \textrm{Hz} and $%
n_{f\left\vert 1\right\rangle }^{UL}=0.2\times 10^{12}$ \textrm{cm}$^{-3}$
at the center $z=0$. In addition, the gradient of the magnetic field is
chosen as $\alpha =21$ \textrm{G/m}. Based on these parameters, $\left\vert
\Delta z_{\left\vert i\right\rangle }\right\vert /2l_{z}^{res}<<1$, $%
\left\vert \Delta V_{\left\vert i\right\rangle }\right\vert /\mu <<1$, $%
\Delta z_{mol}=0$ and $\Delta V_{mol}/\mu =0$. Thus, one can safely omit $%
\Delta z_{\left\vert i\right\rangle }$, $\Delta V_{\left\vert i\right\rangle
}$, $\Delta z_{mol}$ and $\Delta V_{mol}$.

From the equilibrium condition for the chemical potential given by Eq. (\ref%
{chem-condition}), one gets the following expression for the density
distribution:
\begin{equation}
n_{m}^{BEC}=\frac{\left( \mu _{eff}+\hbar ^{2}/ma^{2}-2V_{ext}\right) m}{%
2\pi \hbar ^{2}a_{m}},  \label{BEC-Density}
\end{equation}%
\begin{equation}
n_{f\left\vert 1\right\rangle }^{BEC}=n_{f\left\vert 2\right\rangle
}^{BEC}\approx \left[ \frac{2m\left( \mu _{eff}/2-V_{ext}\right) }{\hbar
^{2}\left( 6\pi ^{2}\right) ^{2/3}}\right] ^{3/2},
\end{equation}%
\begin{equation}
n_{f\left\vert 1\right\rangle }^{UL}=n_{f\left\vert 2\right\rangle }^{UL}=
\left[ \frac{2m\left( \mu _{eff}/2-V_{ext}\right) }{\left( 1+\beta
_{1}\right) \hbar ^{2}\left( 6\pi ^{2}\right) ^{2/3}}\right] ^{3/2},
\end{equation}%
\begin{equation}
n_{d}^{UL}=\left[ \frac{4m\left( \mu _{eff}-2V_{ext}\right) }{\left( 1+\beta
_{1}\right) \hbar ^{2}\left( 6\pi ^{2}\right) ^{2/3}}\right] ^{3/2},
\end{equation}%
\begin{equation}
n_{f\left\vert 1\right\rangle }^{BCS}=n_{f\left\vert 2\right\rangle }^{BCS}=
\left[ \frac{2m\left( \mu _{eff}/2-V_{ext}\right) }{\hbar ^{2}\left( 6\pi
^{2}\right) ^{2/3}}\right] ^{3/2},  \label{BCs-density}
\end{equation}%
where $V_{ext}$ takes the form
\begin{equation}
V_{ext}=\frac{1}{2}m\omega _{r}^{2}\left( x^{2}+y^{2}\right) +\frac{1}{2}%
m\omega _{z}^{2}z^{2}.
\end{equation}%
To give a concise presentation, we have introduced
\begin{equation}
\mu _{eff}=2\left( 1+\beta _{1}\right) \frac{\hbar ^{2}\left( 6\pi
^{2}\right) ^{2/3}}{2m}\left( n_{f\left\vert 1\right\rangle }^{UL}\left(
z=0\right) \right) ^{2/3}.
\end{equation}

\begin{figure}[tbp]
\includegraphics[width=0.7\linewidth,angle=270]{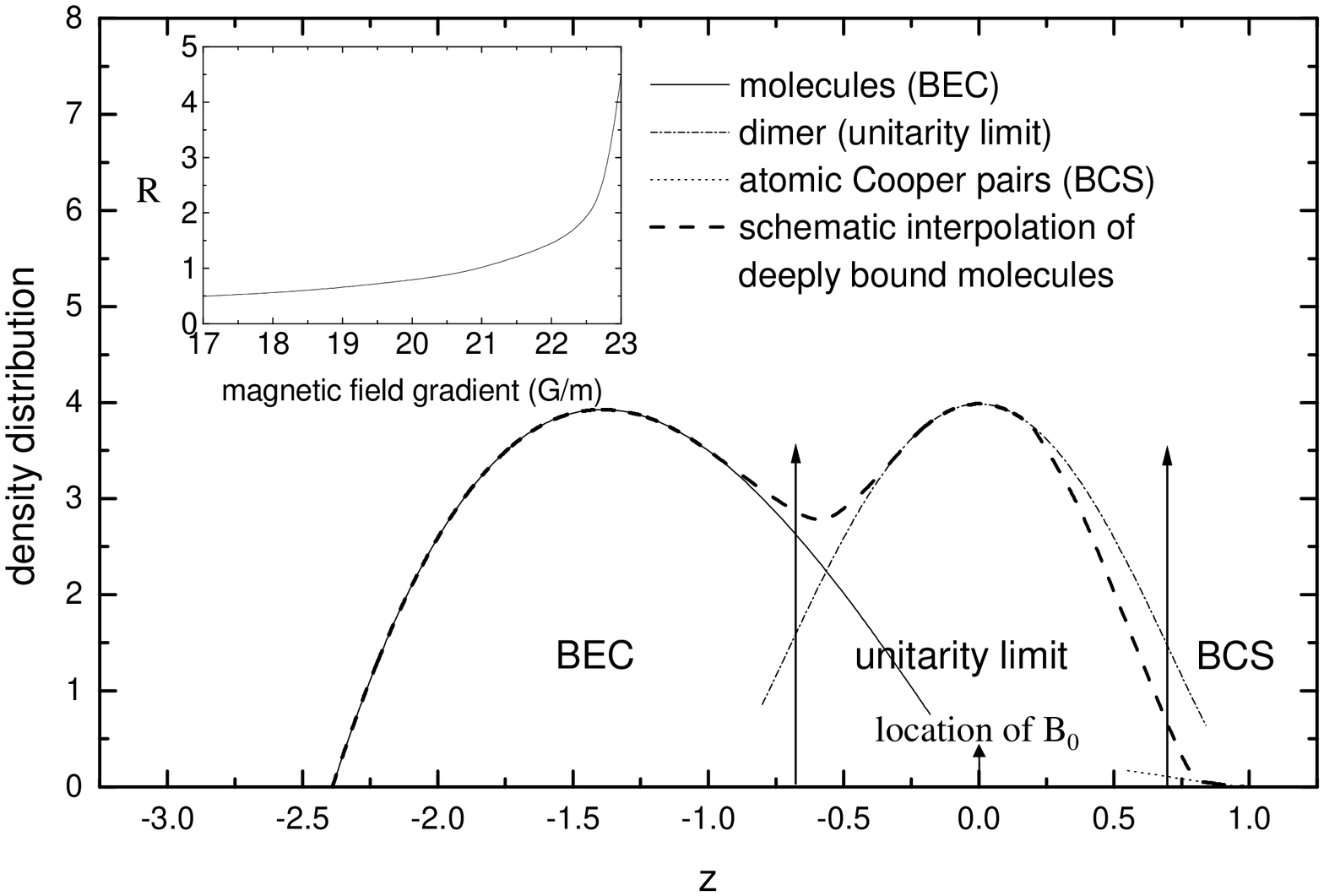}
\caption{Shown is the density distribution of the gases when a
gradient magnetic field of $21$ \textrm{G/m} is used to tune the
atomic scattering length. After the molecules in the regime of
BEC, the dimers in the regime of unitarity limit, and the atomic
Cooper pairs in the regime of BCS are converted into deeply bound
molecules by decreasing non-adiabatically the magnetic field, the
thick dashed line illustrates the density distribution of the
deeply bound molecules. We see that there is a double peak for the
density distribution of the deeply bound molecules which can give
us a clear evidence for an experiment to make the system in the
coexistence of BEC, BCS and unitarity limit. The ratio of the peak
density to valley density is
estimated as $1.4$. In this figure, the density distribution is in unit of $%
n_{f\left\vert 1\right\rangle }^{UL}\left( z=0\right) $, while the
coordinate $z$ is in unit of $R_{z}=\sqrt{\mu _{eff}/m\omega
_{z}^{2}}$. In the inset of this figure, shown is the ratio $R$ of
the peak density due to molecular BEC and ultracold gases in the
unitarity limit for different magnetic field gradient.}
\end{figure}

Using the parameters in this paper, shown in Fig.1 is the density
distribution of the gases where the density is in unit of $n_{f\left\vert
1\right\rangle }^{UL}\left( z=0\right) $, while the coordinate $z$ is in
unit of $R_{z}=\sqrt{\mu _{eff}/m\omega _{z}^{2}}$. The arrow shows the
location $\pm 0.7R_{z}$ which is determined by $\left\vert a\left( z\right)
\right\vert n^{1/3}=1$. The solid line in Fig.1 shows the density
distribution of molecular BEC $n_{m}^{BEC}$ ($z<-0.7$), while the dot-dashed
line in this figure shows the density distribution $n_{d}^{UL}$ ($-0.7<z<0.7$%
) of the dimers in the unitarity limit. The dotted line shows the density
distribution of $n_{f\left\vert 1\right\rangle }^{BCS}$ ($z>0.7$).
Analogously to the experiments in \cite{JIN-fermion,MIT-fermion}, we
consider the process that the magnetic field is decreased non-adiabatically
so that the fermionic atom pairs (i.e. molecules on the BEC side, dimers in
the unitarity limit, and atomic Cooper pairs on the BCS side before the
decreasing of the magnetic field) are converted into deeply bound molecules.
After this non-adiabatical process, from the density distribution of
molecular BEC (solid line) and dimers in the unitarity limit (dot-dashed
line), we see that the coexistence of the gases in the form of BCS, BEC and
unitarity limit can be clearly shown through the double peaks in the density
distribution of the deeply bound molecules. The thick dashed line shows a
schematic interpolation of the density distribution of the deeply bound
molecules. The evidence for the coexistence of the regimes of BCS, BEC, and
unitarity limit can be also shown through the non-symmetric density
distribution of the deeply bound molecules about $z=0$ after the
non-adiabatic process. The inset in the figure shows the ratio $R$ between
the peak density of the regime of BEC and unitarity limit for different
magnetic field gradient. Through the density distribution for different
magnetic field gradient, our researches show that for $18$ \textrm{G/m}$%
<\alpha <22$ \textrm{G/m}, there is obvious double peak density distribution.

In summary, we show that for $^{6}Li$ at the narrow resonance $B_{0}=543.25$
\textrm{G}, by using a gradient magnetic field to change the scattering
length, one can make the gases become the coexistence of the regimes of BCS,
BEC and unitarity limit. In the case of thermal and chemical equilibrium, it
is shown that with appropriate parameters there is a double peak in the
density distribution of the deeply bound molecules after the non-adiabatic
decreasing of the magnetic field. This can give us a clear evidence for the
coexistence of the regimes of BCS, BEC and unitarity limit in an experiment.

We acknowledge the useful discussions with Dr. Cheng Chin. This work is
supported by NSFC under Grant Nos. 10205011, 10474117 and 10474119, NBRPC
under Grant Nos. 2005CB724508 and 2001CB309309.


\end{document}